\documentclass[aip,
 sd,
 amsmath,amssymb,
 reprint,
]{revtex4-1}

\usepackage{graphicx}
\usepackage{dcolumn}
\usepackage{bm}
\usepackage{mathtools}
\usepackage{xcolor}

\DeclarePairedDelimiter\floor{\lfloor}{\rfloor}

\begin{document}

\preprint{AIP/123-QED}

\title{Bitcoin market route to maturity? Evidence from return fluctuations, temporal correlations and multiscaling effects.}

\author{Stanis\l aw Dro\.{z}d\.{z}}
\altaffiliation{Author to whom correspondence should be addressed. Electronic address: stanislaw.drozdz@ifj.edu.pl}
\affiliation{Complex Systems Theory Department, Institute of Nuclear Physics Polish Academy of Sciences, ul. Radzikowskiego 152, 31--342 Krak\'ow, Poland.}
\affiliation{Faculty of Physics, Mathematics and Computer Science, Cracow University of Technology, ul. Warszawska 24, 31--155 Krak\'ow, Poland.  }
\author{Robert G\c{e}barowski}
\affiliation{Faculty of Physics, Mathematics and Computer Science, Cracow University of Technology, ul. Warszawska 24, 31--155 Krak\'ow, Poland.  }
\author{Ludovico Minati}
\affiliation{Complex Systems Theory Department, Institute of Nuclear Physics Polish Academy of Sciences, ul. Radzikowskiego 152, 31--342 Krak\'ow, Poland.}
\author{Pawe\l{} O\'swi\c{e}cimka}
\affiliation{Complex Systems Theory Department, Institute of Nuclear Physics Polish Academy of Sciences, ul. Radzikowskiego 152, 31--342 Krak\'ow, Poland.}
\author{Marcin W\c{a}torek }
\affiliation{Complex Systems Theory Department, Institute of Nuclear Physics Polish Academy of Sciences, ul. Radzikowskiego 152, 31--342 Krak\'ow, Poland.}
         
\date{\today}

\begin{abstract}
Based on 1-minute price changes recorded since year 2012, the fluctuation properties of the rapidly-emerging Bitcoin (BTC) market are assessed over chosen sub-periods, in terms of return distributions, volatility autocorrelation, Hurst exponents and multiscaling effects. The findings are compared to the stylized facts of mature world markets. While early trading was affected by system-specific irregularities, it is found that over the months preceding Apr 2018 all these statistical indicators approach the features hallmarking maturity. This can be taken as an indication that the Bitcoin market, and possibly other cryptocurrencies, carry concrete potential of imminently becoming a regular market, alternative to the foreign exchange (Forex). Since high-frequency price data are available since the beginning of trading, the Bitcoin offers a unique window into the statistical characteristics of a market maturation trajectory.
\end{abstract}

                            
\maketitle

\begin{quotation}
Throughout history, the emergence of money constitutes one of the most significant achievements of civilizations, which simultaneously conditions their further evolution. By means of a process analogous to spontaneous symmetry-breaking in physical systems, money acquires the status of a universally-demanded commodity, and with it an inherent ability to serve as medium of exchange. In the contemporary high-technology era, information can readily become a high-value commodity, and the recent invention of the Bitcoin as a candidate currency naturally reflects this fact. A question that urgently needs addressing is whether, and to what extent, the emerging Bitcoin market exhibits the stylized statistical features which universally characterize mature global markets. Such question is important both theoretically, in reference to our understanding of the emergence of money, and practically, with immediate implications for policy-making. Here, it is demonstrated that over the recent months the Bitcoin market has become truly indistinguishable from mature markets according to the most important complexity characteristics, related to the return distribution, temporal correlations and multi-scaling effects, even including their generalization to discrete scale invariance, which manifests itself through log-periodic oscillations accompanying the large-scale trend reversals.
\end{quotation}

\section{Introduction}
Money is the most liquid economic asset: it serves as a universal medium of exchange, with an inherent ability to buy, store wealth and act as unit of account. The origin of money and what precisely makes moneyness out of different economic assets and commodities remain an active area of interdisciplinary research~\cite{menger1892,bak2001,lapavitsas2005}. Capturing the sheer complexity of the collective dynamics at play in the global foreign exchange (Forex) market represents a substantial challenge for traders, economists and physicists alike \cite{ghashghaie1996,drozdz2010,kwapien2012}. In recent years, arguably the most important advances in this area have come from agent--based computational economics models~\cite{yasutomi1995,yasutomi2003,samanidou2007,gorski2010,oswiecimka2015}.\\
The commodity-based currency is a nearly extinct concept nowadays, with fiat currency relentlessly propelling financial markets around the clock. Over the last decades, the advent of electronic banking and trading, and eventually algorithmic trading, has drastically increased the frequency and security of trading, weaving complex relationships between information/events and fluctuations, and furthermore supporting the development of more and more abstracted derivative products. In 2008, Satoshi Nakamoto -- a pseudonym for the anonymous inventor(s) whose identity remains unconfirmed -- put forward a cryptographic protocol allowing peer--to--peer payments without need for endorsement or endowment by any third--party entity. Its monetary unit, known as the Bitcoin (BTC), accordingly needs not be issued by any central bank or government~\cite{berentsen2018}. This disruptively new concept has become a topic of substantial social and scientific interest, even concern, with potentially deep implications for market regulation and stability~\cite{kristoufek2013}. At the time of writing, more than a thousand electronic currency and transaction protocols using classical asymmetric cryptography technology have emerged, including for example the Ethereum, Ripple, Bitcoin Cash and Litecoin. Moreover, the currently huge volatility of the BTC market, together with its large capitalization, is very likely to spearhead the emergence of other comparably strong cryptocurrencies, with potentially substantial impact on the global trading system.\\
The most informative measures of any market dynamics are derived from the statistical features of the corresponding price fluctuations $P(t)$. These include return $R_{\Delta t}=\log(P(t+\Delta t))-\log(P(t))$ distributions for a given time-lag $\Delta t$, temporal dependences of unsigned return distributions reflecting volatility correlations, the market's susceptibility to persistence as quantified by the Hurst exponents, as well as more subtle non-linear correlations which may manifest as scale-free relationships~\cite{kwapien2012}. Here, a comprehensive account of the emerging BTC market from these perspectives is provided, and references to the notion of ``mature market'' are understood in an econophysics sense~\cite{kertesz1999,cont2001,zheng2014}. Namely, they refer to the statistical features that are empirically-identified and ubiquitous in all well-established, large-capitalization and high-liquidity world markets. These features include: i) fat-tailed return $R_{\Delta t}$ distribution\cite{lux1996,dacorogna2001}, which for sufficiently small $\Delta t$ is to a good approximation scale-free with scaling exponent $\gamma \approx 3$~\cite{gopi1998,gopi1999}, is therefore known as the inverse cubic power-law~\cite{gabaix2003}, ii) temporal autocorrelation of returns $R_{\Delta t}$ that almost instantly decays to zero~\cite{fama1965,gopi1999,drozdz2010,kwapien2012}, iii) volatility autocorrelation which, by contrast, remains positive over long periods of time~\cite{gopi1999,mantegna2000}, iv) Hurst exponent fluctuating close to 0.5~\cite{grech2004,matteo2005,cajueiro2006,tabak2006}, and v) multiscaling effects~\cite{halsey1986} in the temporal organization of returns, which manifest as broad singularity spectra~\cite{oswiecimka2005,drozdz2010,oh2012,shahzad2017}, reflecting the presence of nonlinear correlations in this organization~\cite{calvet2002,kwapien2012}.
 
\section{Bitcoin returns and volatility}
\begin{figure}
\includegraphics[scale=0.29]{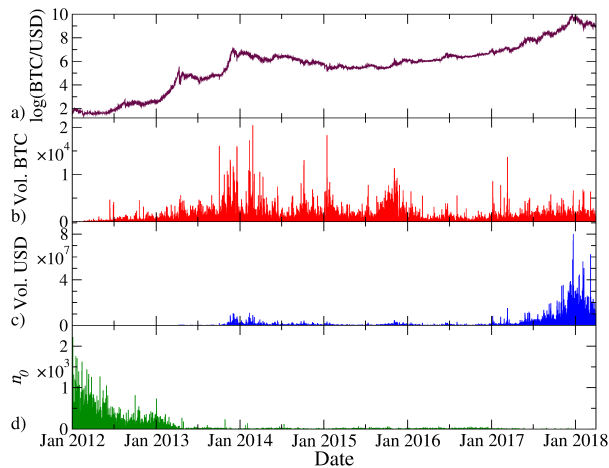} 
\caption{\label{fig:BTCdata}Long-term trends. a) Logarithm of the Bitcoin price expressed in US Dollar (BTC/USD). b) Exchange volume expressed in terms of number of BTC exchanged per 1-hour bin. c) Exchange volume expressed in terms of USD per 1-hour bin. d) Number of consecutive 1-minute bins with zero returns.}
\end{figure}

Publicly-available BTC trading data were downloaded from the Bitstamp exchange (Luxembourg)~\cite{data2018}, and span the period from Jan 01, 2012 to Mar 31, 2018 with a sampling frequency of 1 minute. The corresponding price changes, volume traded expressed in terms of number of Bitcoins exchanged, BTC value in United States Dollars (USD) and number of consecutive 1-minute bins with zero returns are charted in Fig.~\ref{fig:BTCdata}. During the early trading phase, frequent and extended time intervals with no price change, on the order of $10^3$ min, are observed, and plausibly reflect the absence of any transactions. Contrariwise, the most recent data at the time of writing indicate that BTC volume traded expressed in USD is suddenly entering a new regime, characterized by an average increase of about one order of magnitude.

\subsection{Distribution of return fluctuations}
\begin{figure}
\includegraphics[scale=0.29]{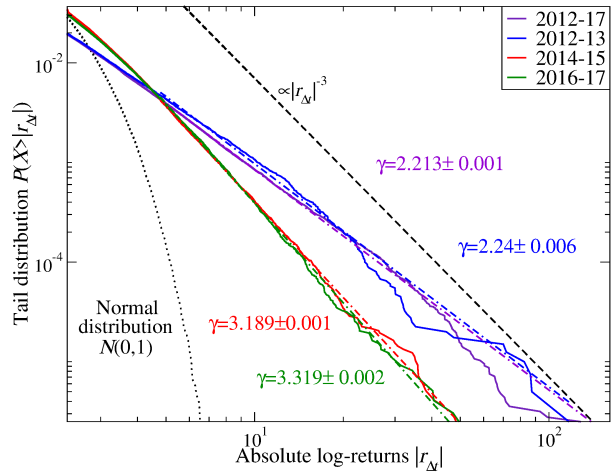}
\caption{\label{fig:Distr}Tail distribution of the absolute normalized log-returns $P(X>|r_{\Delta t})|)$ for the BTC price.}
\end{figure}

It is a well-established fact that, in contemporary mature markets (stock, commodities, Forex), the distribution of return fluctuations on relatively short time-lags $\Delta t$ follows to a good approximation an inverse cubic power-law~\cite{gopi1998,gopi1999,kwapien2012}, i.e. the large-event tails of the return distributions follow $P(X>r_{\Delta t}) \sim r_{\Delta t}^{-\gamma}$ where $r_{\Delta t}=(R_{\Delta t}-\mu)/ \sigma$ denotes the normalized returns ($\mu$, $\sigma$ denote mean and standard deviation) and $\gamma \approx 3$. Such law is understood to reflect positive correlations between large market's movements and the trades of large participants~\cite{gabaix2003,gabaix2006,rak2013}. In the case of the BTC, as shown in Fig.~\ref{fig:Distr} for $\Delta t = 1\textrm{ minute}$, such relationship unequivocally emerges insofar as the early trading phase, i.e. the first two years 2012-2013 in our dataset, is excluded. In these two years, $P(r_{\Delta t})$ develops significantly heavier power-law tails having $\gamma \approx 2.2$. It is noteworthy that these two years dominate the tail thickness, since $\gamma$ remains essentially unchanged when the entire 2012-2017 period is taken into account. The tail thickness approaches the inverse cubic power-law when either of the two more recent biennia is considered: for years 2014-2015 $\gamma \approx 3.2$ is observed, and similarly for years 2016-2017 $\gamma \approx 3.3$ is observed. These values are very close, for example, to those observed over the period from Jan 2, 2018 to Mar 30, 2018, for the rates EUR/USD, GBP/USD and GBP/JPY, for which respectively $\gamma=3.1\pm0.2$, $\gamma=3.1\pm0.2$ and $\gamma=3.2\pm0.1$.

\subsection{Volatility autocorrelation}
\begin{figure}
\includegraphics[scale=0.29]{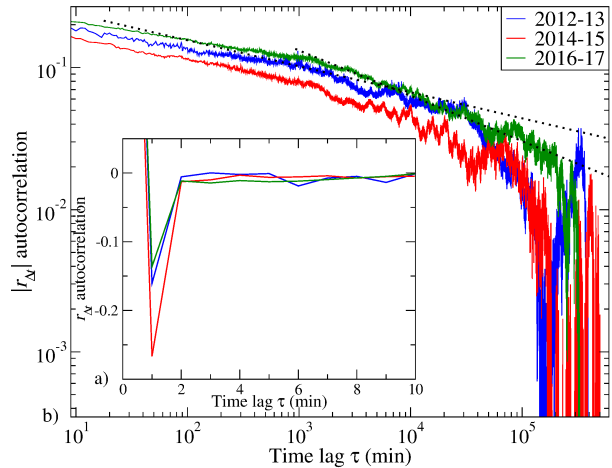}
\caption{\label{fig:BTCauto} Autocorrelation function of a) the normalized log-returns $r_{\Delta t}$ and b) the absolute normalized log-returns (volatility) $|r_{\Delta t}|$ of the BTC price.}
\end{figure}

When applied to returns series $f(t)= r_{\Delta t}(t)$ in the existing markets, the autocorrelation function $c(\tau) = \langle f(t + \tau) f(t) \rangle$, where $\langle...\rangle$ denotes average over $t$, knowingly drops to zero almost immediately~\cite{kwapien2012}, reflecting a very quick disappearance of correlations in the signs of returns. As shown in Fig.~\ref{fig:BTCauto}a, the same applies to the BTC market, which even develops a ``correlation hole'', i.e. $c(\tau)$ drops below zero. Such effect can also be seen in other currencies~\cite{xu2003,drozdz2010} as well as in the stock markets~\cite{fama1965,palagyi1999}, revealing the tendency for returns to change sign between adjacent time-steps.\\
On the other hand, as visible in Fig.~\ref{fig:BTCauto}b the autocorrelation of unsigned returns, $f(t)= \vert r_{\Delta t}(t) \vert$, that is, of volatility, develops long-range, power law-like correlations which extend over an interval on the order of $10^5$ minutes, i.e. two months. This phenomenon originates from the so-called volatility clustering typical of other financial markets, and the sudden fall of $c(\tau)$ for $\vert r_{\Delta t}(t) \vert$ at time horizons beyond $10^5$ minutes reflects the average time-span of the underlying volatility clusters~\cite{drozdz2009}. Attention needs to be drawn to the fact that autocorrelation is strongest and most persistent during the most recent trading period under consideration, i.e. years 2016-2017. As volatility is a nonlinear (i.e., modulus as here) function of returns, the volatility autocorrelation encodes correlations that are nonlinear in time from the perspective of returns. 

\section{Nonlinear temporal correlations}
At present, the most efficient and accurate methods to study diverse temporal correlations, including nonlinear and multifractal correlations~\cite{calvet2002}, in single time-series as well as among multiple time-series, are based on inspecting the scaling properties of the varying-order moments of fluctuations, evaluated after appropriate detrending. The most general of these methods, termed Multifractal Cross-Correlation Analysis (MFCCA), consists of the steps described below~\cite{kantelhardt2002,podobnik2008,zhou2008,oswiecimka2014}. 

\subsection{Methods based on detrending}  
\label{methods}
Given two time-series $x_i$, $y_i$ where $i=1,2...T$, the signal profile is separately calculated for each of them according to
\begin{equation}
X(j) =\sum_{i=1}^j[x_{i}-\langle x\rangle] ,\quad
Y(j) =\sum_{i=1}^j[y_{i}-\langle y\rangle],
\end{equation}
where $\langle \rangle$ denotes averaging over the entire time-series. These signal profiles are subsequently split into $2M_s$ disjoint segments $\nu$ of length $s$ starting both from the beginning and the end of the profile, where $M_s=\floor*{T/s}$ and where in each segment $\nu$, the assumed trend is estimated by fitting a polynomial of order $m$, namely $P^{(m)}_{X,\nu}$ for $X$ and $P^{(m)}_{Y,\nu}$ for $Y$. In typical cases, $m=2$ provides an optimal choice~\cite{oswiecimka2006}. By subtracting this trend from the series, the detrended cross-covariance within each segment is obtained
\begin{multline}
F_{xy}^{2}(\nu,s)=\frac{1}{s}\Sigma_{k=1}^{s}\lbrace
(X((\nu-1)s+k)-P^{(m)}_{X,\nu}(k)) \\ \times
(Y((\nu-1)s+k)-P^{(m)}_{Y,\nu}(k))\rbrace,
\label{Fxy2}
\end{multline}
which can in turn to be used to define the $q$th-order covariance function~\cite{oswiecimka2014}
\begin{equation}
F_{xy}^{q}(s)=\frac{1}{2M_s}\Sigma_{\nu=1}^{2M_s} {\rm
sign}(F_{xy}^{2}(\nu,s))|F_{xy}^{2}(\nu,s)|^{q/2},
\label{Fq}
\end{equation}
where ${\rm sign}(F_{xy}^{2}(\nu,s))$ denotes the sign of $F_{xy}^{2}(\nu,s)$.
Fractal cross-dependencies between given time-series $x_i$ and $y_i$ then manifest themselves in scaling relations of the form
\begin{equation}
F_{xy}^{q}(s)^{1/q}=F_{xy}(q,s) \sim s^{\lambda_q}, 
\label{Fxy}
\end{equation}
where $q\neq0$ and $\lambda_q$ is the corresponding scaling exponent, whose dependence on $q$ reflects a richer, multifractal character of correlations in the time-series as compared to the monofractal case, for which $\lambda_q$ is $q$-independent.\\
The conventional MFDFA procedure~\cite{kantelhardt2002} of calculating the singularity spectra for single time-series is a special case of the above MFCCA procedure, wherein $x_i=y_i$. The Eq.~(\ref{Fq}) then reduces to
\begin{equation}
F(q,s)=\Big[\frac{1}{2M_s}\sum^{2M_s}_{\nu=1}{[F^2(\nu,s)]^{\frac{q}{2}}}\Big]^{\frac{1}{q}},
\label{F}
\end{equation}
and, as in Eq.~(\ref{Fxy}), multifractality and monofractality are reflected in
\begin{equation}
F(q,s) \sim s^{h(q)},
\label{Hq}
\end{equation}
where $h(q)$ denotes the generalized Hurst exponent, and $h(2)$ is its ordinary case. The singularity spectrum $f(\alpha)$, also referred to as the multifractal spectrum, can thereafter be computed from the following relations
\begin{equation}
\alpha=h(q)+qh'(q), \quad f(\alpha)=q[\alpha-h(q)]+1,
\end{equation}
where $\alpha$ denotes the H\"older exponent characterizing the singularity strength, and $f(\alpha)$ reflects the fractal dimension of support of the set of data points whose H\"older exponent equals $\alpha$.\\
In the case of multifractals, the shape of the singularity spectrum typically resembles an inverted parabola; furthermore, the degree of complexity is straightforwardly quantified by the width of $f(\alpha)$, simply defined as $\Delta \alpha = \alpha _\textrm{max}  - \alpha _\textrm{min}$, where $\alpha _\textrm{min}$ and $\alpha _\textrm{max}$ correspond to the opposite ends of the $\alpha$ values as projected out by different $q$-moments (Eq.~(\ref{F})).  Another important feature of $f(\alpha)$ is its asymmetry (skewness), which can be quantified by the asymmetry index~\cite{drozdz2015} $A_{\alpha}=\frac{\Delta \alpha _L - \Delta \alpha _R}{\Delta \alpha _ L + \Delta \alpha _R}$ where $\Delta \alpha _L = \alpha _0 - \alpha _\textrm{min}$ and $\Delta \alpha _R = \alpha _\textrm{max} - \alpha _0$, and $\alpha _0$ corresponds to the maximum $f(\alpha)$ on the spectrum. Positive and negative values of $A _\alpha$ reflect, respectively, left- and right-sided asymmetry of $f(\alpha)$, which correspond to more developed multifractality on the level of large or small fluctuations in the time-series.\\
Lastly, a family of the fluctuation functions defined by Eq.~(\ref{Fq}) can also be used to introduce a $q$-dependent detrended cross-correlation ($q$DCCA)~\cite{kwapien2015} coefficient
\begin{equation}
\rho_q(s) = \frac{F_{xy}^q(s)}{\sqrt{ F_{xx}^q(s) F_{yy}^q(s) }},
\label{rho.q}
\end{equation}
which allows quantifying the degree of cross-correlation between two time-series $x_i$ and $y_i$ after detrending, over varying time-scales $s$ and, by varying the parameter $q$, identifying the range of detrended fluctuation amplitudes which are most strongly correlated between the two signals~\cite{kwapien2015,kwapien2017}. 

\subsection{Hurst exponent}
\begin{figure}
\includegraphics[scale=0.29]{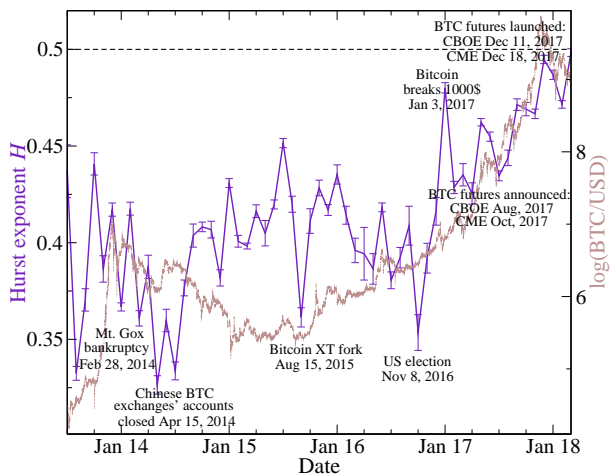}
\caption{\label{fig:Hurst}Hurst exponent $H$ calculated over 1 month windows for the BTC price from Jul 2013 to Mar 2018. Error bars reflect the standard error of regression slope. Highlighted events: 1) One of the biggest Bitcoin exchange - Mt. Gox bankrupted; 2) Chinese BTC exchanges' accounts closed - the People's Bank of China's frequently updated restrictions against Bitcoin finally pressure Chinese banks to issue a deadline against several Bitcoin exchanges, requiring them to close down their accounts by April 15; 3) Bitcoin XT fork - Bitcoin core developers release a separate version of the Bitcoin client software, called Bitcoin XT, culminating fears that the Bitcoin community may not be able to reach a consensus on the issue, and the blockchain may hard fork, resulting in two separate versions of Bitcoin's global ledger; 4) 2016 United States presidential elections; 5) BTC price breaks the psychological ceiling of 1000 USD, mass media coverage brings in influx of new users that plausibly drive price even higher; 6) Futures contract on BTC prices announced; 7) Chicago Board Options Exchange (CBOE) and Chicago Mercantile Exchange (CME) launched futures contract on BTC prices.}
\end{figure}

One of the principal statistical properties of time-series is the Hurst exponent, which quantifies the degree of persistence and as such has particular relevance in the context of financial time-series~\cite{ausloos2000}. Here, the Hurst exponent was estimated as $H=h(q=2)$ following the procedure prescribed by Eqs.~(\ref{F}) and (\ref{Hq}) over 1-month time windows, each one comprising $>\textrm{40,000}$ data points, limiting estimation error, and starting from Jul 2013; earlier windows contain excessive consecutive zero returns, which bias the procedure unacceptably. As charted in Fig.~\ref{fig:Hurst}, during the early trading period the Hurst exponent $H(t)$ for the BTC/USD price takes values $\ll0.5$, hallmarking the strong anti-persistence which is expected for such high-risk emerging market; it thereafter gradually approaches 0.5 from below, recently becoming very close to this value, which is interpreted as a strong indication that the market is approaching maturity~\cite{matteo2003}. By comparison, over the period from Jan 2, 2018 to Mar 30, 2018, for the rates EUR/USD, GBP/USD and GBP/JPY, one observes respectively $H=0.48\pm0.001$, $H=0.49\pm0.001$ and $H=0.50\pm0.001$. It is noteworthy that, during the maturation process, the Hurst exponent of the BTC dropped sharply as a consequence of events apparently causing uncertainty, in particular the bankruptcy of the Mt. Gox Bitcoin exchange, following it the closure of the Chinese BTC exchange accounts accompanied by related regulatory changes~\cite{kristoufek2015}, and finally a fork in the Bitcoin software and the 2016 United States presidential elections. On the other hand, events which appear to be correlated positively with the maturation process include the successful breaking of a psychological $1,000$ USD barrier in early 2017, followed by the announcement and subsequently launch of BTC future contracts in fall of that year.  

\subsection{Multiscaling}
\begin{figure}
\includegraphics[scale=0.29]{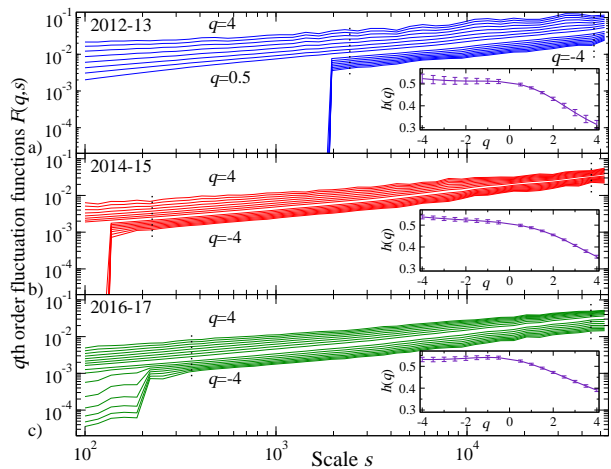}
\caption{\label{fig:Fq}Multifractal analysis of the BTC price in all trading biennia. Family of the $q$th-order fluctuation functions $F(q,s)$ for $q\in [-4,4]$, and corresponding generalized Hurst exponent $h(q)$ (inserts) evaluated over the range of scales indicated by the dashed vertical lines, in a) years 2012-13, b) years 2014-15 and c) years 2016-17, }
\end{figure}

\begin{figure}
\includegraphics[scale=0.29]{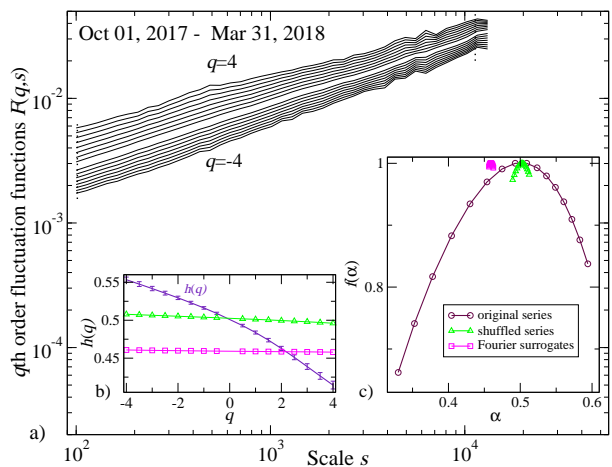}
\caption{\label{fig:Fq17}Multifractal analysis of the BTC price in the most recent semester, i.e. October 01, 2017 to  March 31, 2018. a) Family of the $q$th-order fluctuation functions $F(q,s)$ for $q\in [-4,4]$. b) Generalized Hurst exponent $h(q)$. c) Singularity spectrum $f(\alpha)$. b) and c) were evaluated over the range of scales denoted by the dashed vertical lines. Circles denote the original BTC series, while triangles their randomly shuffled and squares the Fourier phase randomized surrogates.}
\end{figure}

The formalism presented in Section III.A allows studying even more subtle quantitative characteristics related to the nonlinear correlations in time-series, which are encoded in the scale-dependence of the fluctuation functions $F(q,s)$. These function are shown, separately for the BTC return time-series in each of the three two-year time periods and given $q\in [-4,4]$, in Fig.~\ref{fig:Fq}. From the earliest through the most recent trading period, the quality of scaling systematically improves, and the range of scales over which scaling-like behavior of $F(q,s)$ is observed extends. The $q$-dependence of the corresponding generalized Hurst exponents $h(q)$ estimated from the chosen scaling range for each trading period, denoted by the dashed vertical lines, is shown in the corresponding panels within Fig.~\ref{fig:Fq}.\\
The most relevant behavior of the fluctuation functions $F(q,s)$ pertains to the negative values of $q$, which filter out the small fluctuations. As shown in Fig.~\ref{fig:Fq}a, due to number of consecutive 1-minute bins with zero returns in the earliest trading biennium (years 2012-2013) the fluctuation functions even become singular up to the scales $s>10^3$. Only after that point scaling becomes detectable, but the corresponding $h(q)$ function shows essentially no $q$-dependence. In the intermediate biennium (years 2014-2015), the singularity regime shrinks by about one order of magnitude, and in the most recent biennium (years 2016-2017), it largely disappears, with only some residual distortions originating from trading around the time of the US presidential election. Nevertheless, multifractal scaling does not emerge, even in the regime of satisfactory scaling.\\
In striking contrast, fully developed multifractality can be readily appreciated for the most recent semester, i.e. Oct 1, 2017 to Mar 31, 2018. The corresponding fluctuation functions $F(q,s)$ and the resulting multifractal spectrum are shown in Fig.~\ref{fig:Fq17} (circles). In all respects, these results resemble remarkably closely those obtained for mature markets~\cite{kwapien2012}, and in particular for the Forex~\cite{drozdz2010}. Namely, the scaling regime comprises the whole span of scales $s$ considered before, and multifractality is convincingly reflected both in the $q$-dependence of $h(q)$ and in the singularity spectrum $f(\alpha)$, which has a somewhat left-sided asymmetry ($A _\alpha\approx0.33$). The latter is almost universally characteristic of mature markets, and can be explained by the fact that small price changes, as filtered out by the negative $q$-values, are by their nature more noisy and their hierarchical organization is poorer~\cite{drozdz2015}. Importantly, all these multiscaling effects, as it is also shown in Fig.~\ref{fig:Fq17}, completely disappear on surrogate time-series~\cite{schreiber2000}, obtained from the original BTC return time-series by shuffling (squares), or by Fourier phase randomization (triangles). Since the latter preserves the linear correlations, the location of the corresponding $f(\alpha)$ reflects the average Hurst exponent characteristic to the period considered. In the randomly shuffled series all correlations are removed, and consequently $f(\alpha)$ is centered exactly at 0.5. The non-zero width seen in $f(\alpha)$ in this case is an artifact of the finite length of the time-series considered, and is a known effect~\cite{drozdz2009} for the uncorrelated fat-tailed series.

\subsection{Volatility vs. volume traded}
\begin{figure}
\includegraphics[scale=0.29]{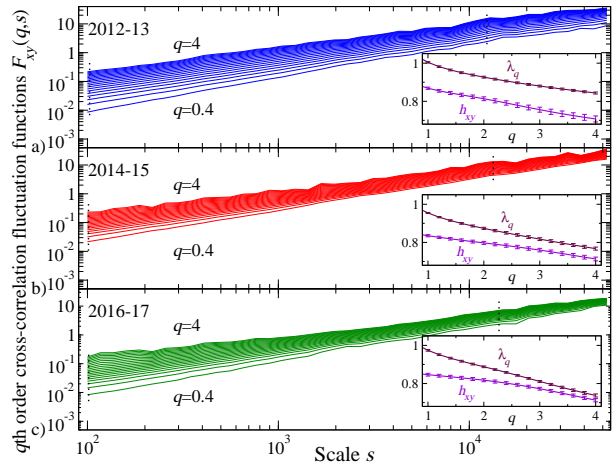}
\caption{\label{fig:Fqcc}Multifractal analysis of BTC volatility $|r_{\Delta t}|$ vs. BTC volume in all trading biennia. Family of the $q$th order cross-correlation fluctuation functions $F_{xy}(q,s)$ for $q\in [0.4,4]$, and corresponding cross-correlation scaling exponents $\lambda_{q}$ and average generalized Hurst exponents $h_{xy}(q)$ (inserts), estimated for $q\in [1,4]$ and over the range of scales indicated by the dashed vertical lines in a) years 2012-13, b) years 2014-15 and c) years 2016-17. Only the $F_{xy}(q,s)$ functions $(q \ge 0.4)$ for which at least an approximate scaling holds are shown.}
\end{figure}

\begin{figure}
\includegraphics[scale=0.29]{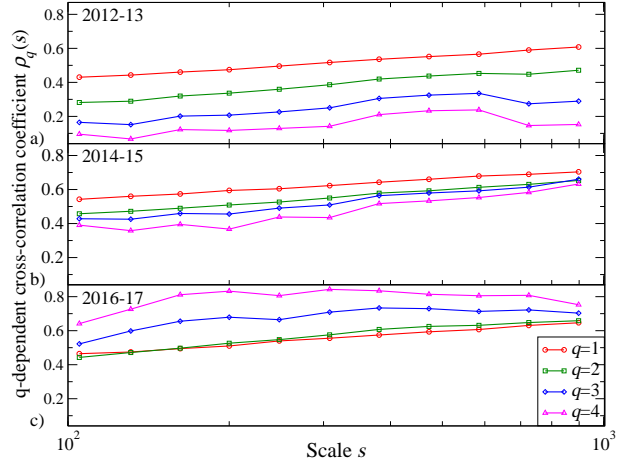}
\caption{\label{fig:pq}The $q$-dependent detrended cross-correlation coefficient $\rho_q$ between BTC volatility $|r_{\Delta t}|$ and Bitcoin volume in all trading biennia as a function of temporal scale $s$ for $q=1,2,3,4$ in a) years 2012-13, b) years 2014-15 and c) years 2016-17.}
\end{figure}

Large fluctuations in mature markets are typically characterized by cross-correlations, even multifractal ones, between volatility and the volume traded~\cite{rak2015}. Within a novel, more advanced approach which is able to detect the degree of synchrony between the two multifractal time-series $x_i$ and $y_i$, the corresponding measures are quantified in terms of the cross-correlation fluctuation functions of Eq.~(\ref{Fxy})~\cite{oswiecimka2014}. When these $F_{xy}(q,s)$ functions develop scaling, the corresponding scaling exponent $\lambda_{q}$ can be determined. Their closeness as regards the average generalized Hurst exponents $h_{xy}(q)=(h_x(q)+h_y(q))/2$ reflects the degree of cross-correlation. Maximal multifractal synchrony corresponds to $\lambda_{q}=h_{xy}(q)$~\cite{oswiecimka2014}. The development of such cross-correlations between the volume traded $(x_i)$ and volatility $(y_i)$ for the BTC market is shown, for each biennium, in Fig.~\ref{fig:Fqcc}. The cross-correlation is identified and reaches maximum synchrony during the most recent period, at large fluctuations ($q$ between 3 and 4). For $q<0.4$, which probes smaller fluctuations, the $F_{xy}(q,s)$ functions in the present case fluctuate between positive and negative values, thus exhibit no scaling and therefore are not shown in Fig.~\ref{fig:Fqcc}. This indicates the entire disappearance of cross-correlations on the level of smaller fluctuations. Notably, no such cross-correlation effects, nor even trace of scaling in $F_{xy}(q,s)$ for any $q$-value, is observed when volatility is replaced by returns.\\
A complementary view on this aspect of the BTC market is offered by the $\rho_q(s)$ coefficient (Eq.~(\ref{rho.q})). The result of its application to the present BTC volume traded and volatility time-series is shown in Fig.~\ref{fig:pq} for time-scales $s$ between $10^2$ and $10^3$ min. While in the first biennium the correlations are relatively weak, especially on the level of large fluctuations ($q=4$), they gradually become much stronger, and in the most recent period the situation becomes opposite, as the large fluctuation cross-correlations dominate.\\
In view of the theoretical concepts supporting the validity of the inverse cubic power-law, demonstrated for the BTC in Section II A, and based on correlations between the large market's movements and the trades of large participants~\cite{gabaix2003,gabaix2006}, these observations provide further arguments for the onset of maturity characteristics in the BTC market.

\subsection{Log-periodic Bitcoin price patterns}
\begin{figure}
\includegraphics[scale=0.29]{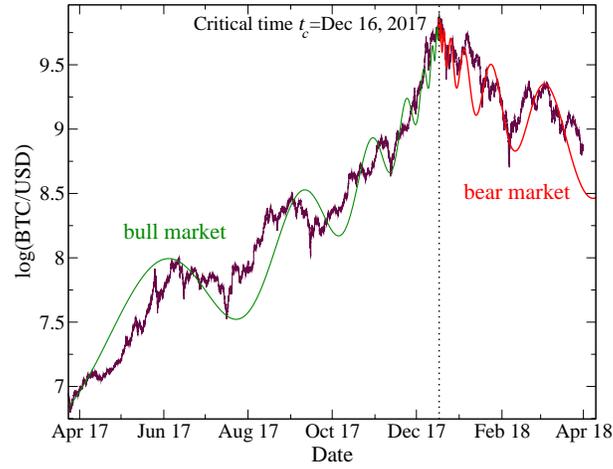}
\caption{\label{fig:LPPL} Logarithm of the BTC price and log-periodic fits: i) accelerated, bubble phase, bull market (green, $A=10.2$, $B=-0.32$, $C=0.05$, $m=0.4$, $\lambda=2$, $\phi=2.12$) and ii) decelerated, anti-bubble phase, bear market (red, $A=10.2$, $B=-0.17$, $C=0.06$, $m=0.4$, $\lambda=2$, $\phi=-1.84$). Spearman rank-order correlation coefficient between the data and fit $\rho=0.96$.}
\end{figure}

As shown in Fig.~\ref{fig:LPPL}, mid-way through the most recent trading semester, namely on Dec 16, 2017, the Bitcoin (BTC) market experienced a spectacular trend reversal, from strongly increasing to sharply decreasing, both large trends being accompanied by well-evident, smaller-scale oscillations. Such phenomena are a well-known product of the dynamics of mature financial markets, and unavoidably emerge in all of them, including stock exchange, commodities, Forex and even the bond markets~\cite{sornette2002,drozdz2003,kwapien2012}. They are, in fact,  analogous to critical phenomena which accompany second-order phase transitions in other physical systems, and closely related to scale invariance. In this case, the scale invariance is however not continuous, but discrete, with a preferred scaling factor $\lambda$, which overlapped to the $\Lambda$-shaped ordinary phase transition introduces log-periodic oscillations. These accumulate exactly at the critical time $t_c$ which indicates, in the financial context, the instant of trend reversal.\\
One of the simplest suitable parameterizations for the temporal evolution of price $P(t)$ is
\begin{equation}
\log(P(t))=A+B(t_c-t)^m+C(t_c-t)^m\cos(\omega\log(t_c-t)-\phi)
\label{log-per}
\end{equation}
for the bubble (increasing) phase, where $\omega = 2\pi$/$\log(\lambda)$. The anti-bubble (decreasing) phase corresponds to changing sign of $t_c-t$. There is some empirical evidence that in contemporary financial markets $\lambda \approx 2$ is common, possibly universal~\cite{drozdz2003,bartolozzi2005}. Assuming therefore $\lambda=2$, and adjusting the remaining parameters in Eq.~(\ref{log-per}), one obtains the fits shown in Fig.~\ref{fig:LPPL}, for $t_c$ corresponding to Dec 16, 2017. These reproduce remarkably well the observed oscillatory patterns, providing additional evidence that the BTC market may have entered a phase of maturity. Activity prior to Apr 2017 was not considered, because the market was not yet mature prior to that point, as indicated by all other markers e.g. $H<0.5$, and as such was not expected to obey a log-periodic relationship.

\section{Commentary}
The disruptiveness of cryptocurrencies, in particular of the Bitcoin (BTC), is well-evident in the striking, possibly unprecedented, rapidity of market capitalization. Around the end of year 2017, namely approximately 8 years after the ``Bitcoin Pizza Day'' which celebrated the time at which 10,000 BTC were worth approximately two pizzas, the same number of BTCs was valued around 20 million USD. At present, the cryptocurrency market as a whole exceeds, in both transaction volume and capitalization, many traditional and well-established financial markets.\\
The present study shows a perhaps more subtle, but not less impacting fact: that, in spite of its virtual nature and novelty, the Bitcoin market has recently and rapidly developed the statistical hallmarks which are empirically observed for all ``mature'' markets like stocks, commodities or Forex. It appears plausible that other cryptocurrencies will follow the same trajectory. This may lead to the emergence of a completely new market, analogous to the the global foreign exchange (Forex) market, wherein cryptocurrencies could hypothetically be traded even in a self-contained manner. In the authors' view, the substantial implications of such prospect for market stability and regulation motivate very urgent research.\\
The Bitcoin phenomenon is also unique in that, given the availability of high-frequency trading data since its inception and the rapidity of evolution (and capitalization), it provides new windows for exploring rigorously the trajectory towards maturity of a currency market. This will in particular motivate further investigations of the entire cryptocurrency market.

\end{document}